\documentclass[prd,twocolumn,tightenlines,showpacs,nofootinbib,amsfonts,amssymb,amsmath]{revtex4}
\newcommand{\etal}{{\it et al.}}
\newcommand{\dd}{{\rm d}}

\begin{document}

\title{Time drift of cosmological redshifts as a test of the Copernican principle}

\author{Jean-Philippe Uzan}
 \email{uzan@iap.fr}
 \affiliation{
             Institut d'Astrophysique de Paris,
             Universit\'e Pierre~\&~Marie Curie - Paris VI,
             CNRS-UMR 7095, 98 bis, Bd Arago, 75014 Paris, France,}

\author{Chris Clarkson}
 \email{chris.clarkson@uct.ac.za}
 \affiliation{Cosmology \& Gravity Group, Department of Mathematics and Applied Mathematics,
 University of Cape Town, Rondebosch 7701, South Africa}

\author{George F.R. Ellis}
 \email{george.ellis@uct.ac.za}
 \affiliation{Cosmology \& Gravity Group, Department of Mathematics and Applied Mathematics,
 University of Cape Town, Rondebosch 7701, South Africa}

\begin{abstract}
We present the time drift of the cosmological redshift in a
general spherically symmetric spacetime. We demonstrate that its
observation would allow us to test the Copernican principle and so
determine if our universe is radially inhomogeneous, an important
issue in our understanding of dark energy. In particular, when
combined with distance data, this extra observable allows one to
fully reconstruct the geometry of a spacetime describing a
spherically symmetric under-dense region around us, purely from
background observations.
\end{abstract}
 \date{28 December 2007}
 \maketitle

\section{Introduction}

\vskip-.35cm Cosmological data is usually interpreted under the
assumption that the universe is spatially homogeneous and
isotropic. This is justified by the Copernican principle, stating
that we are not located at a favoured position in space. Combined
with the observed isotropy, this leads to a Robertson-Walker (RW)
geometry~\cite{Cprinciple}, at least on the scale of the
observable universe.

This implies that the spacetime metric reduces to a single
function of the cosmic time, the scale factor $a(t)$. This
function can be Taylor expanded as $a(t)=a_0+H_0(t-t_0) -
\frac{1}{2}q_0H_0^2(t-t_0)^2+\ldots$ where $H_0$ is the Hubble
parameter and $q_0$ the deceleration parameter. Low redshift
observations~\cite{refsn} combined with the assumption of almost
flatness of the spatial sections, justified mainly by the cosmic
microwave background data~\cite{refdata,uke}, lead to the
conclusion that $q_0<0$: the expansion is accelerating. This
conclusion involves no hypothesis about the theory of gravity or
the matter content of the universe~\cite{refde}, as long as the
Copernican principle holds. This has stimulated a growing interest
in possible explanations~\cite{refde,refde2}, ranging from new
matter fields dominating the dynamics at late times to
modifications of general relativity.

While many tests of general relativity on astrophysical scales
have been designed~\cite{modifGR}, the verification of the
Copernican principle has attracted little attention, despite the
fact that relaxing this assumption may be the most conservative
way, from a theoretical perspective, of explaining the recent
dynamics of the universe without introducing new physical degrees
of freedom~\cite{nature}.

This possibility that we may be living close to the center
(because isotropy around us seems well established
observationally) of a large under-dense region has attracted
considerable interest. In particular, the low redshift
(background) observations such as the magnitude-redshift relation
can be matched~\cite{ltb1} by a non-homogeneous spacetime of the
Lema\^{\i}tre-Tolman-Bondi (LTB) family (that is, spherically
symmetric solution of Einstein equations sourced by pressureless
matter and no cosmological constant). Unfortunately, this simple
extension of the RW universes depends on two free functions (see
below for details) so that the reconstruction is under-determined
and one must fix one function by hand. Thus, one needs at least
one extra independent observation to reconstruct the geometry of
an LTB universe. A limitation to this reconstruction arises
because most data lie on our past light cone. This takes us back
to the observational cosmology program~\cite{ellis85} and the
question~\cite{KS66} of how to extract as much information as
possible about our spacetime from cosmological data alone. Among
many results, it was demonstrated~\cite{LTBy} that the two free
functions of a LTB spacetime can be reconstructed from the angular
distance and number counts, even though evolution effects make it
impossible to be conclusive~\cite{mus}.

Recently, two new ideas were proposed. First, it was
realised~\cite{testCP1} that the distortion of the Planck spectrum
of the CMB allows one to test the Copernican principle. Second, a
consistency relation between distances on the null cone and Hubble
rate measurements in RW universes was derived~\cite{testCP3},
based on the fact that the curvature is constant; this also serves
as an observational test of the Copernican principle.

In this letter, we reconsider the time drift of cosmological
redshift in spacetimes with less symmetries than the RW universe
and we demonstrate how, when combined with distance data, it can
be used to test the Copernican principle, mainly because observing
the thickening of our past light cone brings new information. As
pointed out by Sandage and then McVittie~\cite{Sandage}, one
should expect to observe such a time drift in any expanding
spacetime. This may lead to a better understanding of the physical
origin of the recent acceleration~\cite{Lake07,demodels}, or to
tests of the variation of fundamental constants~\cite{constant}.
This measurement, while challenging, may be achieved with
Extremely Large Telescopes (ELT) and in particular, it is one of
the main science drivers in design of the COsmic Dynamics
EXperiment (CODEX) spectrograph~\cite{Pasquini1}. Our result may
strengthen the scientific case for this project.

We start by introducing observational coordinates, which allow us
to derive the general expression for the time drift in a
spherically symmetric, but not necessarily spatially homogeneous,
universe [see Eq.~(\ref{gen8})]. We show that observation of both
the luminosity distance and the redshift drift allows one to probe
the Copernican principle at low redshifts, when ``dark energy''
dominates [see the consistency relation~(\ref{Crelat})]. We
demonstrate that this expression may be used with distance data to
fully reconstruct the geometry of a LTB spacetime [see
Eq.~(\ref{newcond})].

\section{$\dot z$ in a spherically symmetric universe}

\vskip-.3cm {\bf Observational coordinates.} We consider a
spherically symmetric spacetime in observational coordinates
$\{w,y,\vartheta,\varphi\}$, where $w$ labels the past light-cones
of events along the wordline $\mathcal{C}$ of the observer,
assumed to lie at the center so that $w$ is constant on each past
light cone, with $u^a\partial_a w>0$, $u^a$ being the 4-velocity
of the cosmic fluid, $u_a u^a=-1$. $y$ is a comoving radial
distance coordinate specified down the past light cone of an event
$\mathcal{O}$ on $\mathcal{C}$. Many choices are possible such as
the affine parameter down the null geodesics from $\mathcal{O}$,
the area distance, or the redshift; whatever choice we assume on
the past light cone of $\mathcal{O}$, it is specified on other
past light cones through being comoving with the cosmic fluid,
i.e. $y_{,a} u^a=0$. $(\vartheta,\varphi)$ are angular coordinates
based at $\mathcal{C}$ and propagated parallelly along the past
light cone. The metric in these observational coordinates is
\begin{eqnarray}\label{Obsmetric}
 \dd s^2&=&-A^2(w,y)\dd w^2 + 2A(w,y)B(w,y)\dd y\dd w
 \nonumber\\
 && +C^2(w,y)\dd\Omega^2\ ,
\end{eqnarray}
which is clearly spherically symmetric around the worldline
$\mathcal{C}$ defined by $y=0$. The requirement that the 2-spheres
$\{w,y\}={\rm const.}$ behave regularly around $\mathcal{C}$ when
$y\rightarrow0$ implies~\cite{ellis85} that $A(w,y)\rightarrow
A(w,0)\not=0$, $B(w,y)\rightarrow B(w,0)\not=0$ and
$C(w,y)=B(w,0)y+\mathcal{O}(y^2)$.

There remain two coordinate freedoms in possible rescalings of $w$
and $y$. However, once specified on $\mathcal{C}$, $w$  is
determined on the other worldlines by the condition that $\{w={\rm
const.}\}$ are past light-cones of events on $\mathcal{C}$. This
allows us to arbitrarily choose $A(w,0)$. Also, once specified on
one past light cone, $y$ is determined on all the others because
it is a coordinate comoving with the fluid. This allows us to
choose $B(w_0,y)$ for a given value of $w=w_0$.

On each past light cone, the cross-sectional area of a source is
related to the solid angle $\dd\Omega^2$ under which it is
observed by an observer on $\mathcal{C}$ at $w=w_0$  by
$C^2(w_0,y)\dd\Omega^2$. This implies that $C$ is the angular
distance, $D_A$, i.e. $D_A(y) = C(w_0,y)$. The distance duality
relation~\cite{ddr} then implies that the luminosity distance is
given by $D_L(y)=(1+z)^2D_A$. The redshift is given by
\begin{equation}\label{defz}
 1+z= \frac{\left(u_a k^a\right)_{\rm emission}}{\left(u_a
k^a\right)_{\rm observer}}
 = \frac{A(w_0,0)}{A(w_0,y)}\ ,
\end{equation}
where the matter velocity and photon wave-vector are given by
$u^a=A^{-1}\delta^a_w$ and $k^a=(AB)^{-1}\delta^a_y$ respectively.
We deduce that the isotropic expansion rate, defined by $3H=
\nabla_a u^a$, is given by
\begin{equation}\label{defH}
 H(w,y)=\frac{1}{3A}\left[\frac{\partial_w B(w,y)}{B(w,y)}
 +  2\frac{\partial_w C(w,y)}{C(w,y)} \right]\ .
\end{equation}
For the central observer, who sees the universe isotropic, $H$ is
simply the Hubble expansion rate. At small redshifts, $ H(w,y) =
\frac{\partial_w B(w,0)}{B(w,0)A(w,0)} + \mathcal{O}(y)$, so the
Hubble constant is  $H_0=\partial_w B(w_0,0)/B(w_0,0)A(w_0,0)$.

In the particular case of a dust dominated universe, the
acceleration and vorticity vanish and the fluid 4-velocity can be
expressed as the gradient of the proper time along the matter
worldlines: $u_a=-\partial_a t$. Since we also have
$u_a=-A\partial_a w + B\partial_a y$ we deduce that $\dd t= A\dd w
- B\dd y$ so that $A=\partial_w t$ and $B=\partial_y t$. The
surfaces of simultaneity are thus given by $A\dd w=B\dd y$ and we
have the integrability condition $\partial_y A +
\partial_w B= 0$.

The covariant derivative of $u_a$ is therefore of the form
$\nabla_a u_b=H(g_{ab}+u_au_b)+\sigma_{ab}$, where the shear
$\sigma_{ab}$ is symmetric, traceless, and satisfies
$u^a\sigma_{ab}=0$. The scalar shear
$\sigma^2=\sigma^{ab}\sigma_{ab}/2$ is consequently the only
non-vanishing kinematical variable and is given by
\begin{equation}
 \sigma(w,y) =\frac{2}{\sqrt{3}}\frac{1}{A}\left(\frac{\partial_w B}{B} -
 \frac{\partial_w C}{C} \right)\ ,
\end{equation}
where an arbitrary sign has been chosen. The regularity conditions
imply $\sigma(w,0)=0$, which is expected since the expansion is
observed to be isotropic about the central worldline.

{\bf Expression of the redshift drift.} From the
expression~(\ref{defz}), it is straightforward to deduce that
$\dot z \equiv \frac{\delta z}{\delta w}(w_0,y)$ is given by
\begin{eqnarray}\label{defzdot}
 \dot z(w_0,y)
 &=& (1+z)
 \left[\frac{\partial_w A(w_0,0)}{A(w_0,0)} - \frac{\partial_w A(w_0,y)}{A(w_0,y)}
 \right].
\end{eqnarray}
Now, we can choose $w$ such that $A(w_0,0)=1$. Then, on our past
light cone, we can choose $y$ such that $\partial_w\ln
B(w_0,y)=\partial_w\ln A(w_0,y)$. Note however that such a choice
is not possible for all $w$. It follows that
\begin{equation}\label{gen8}
 \dot z(w_0,y) = (1+z)H_0 - H(w_0,y) -\frac{1}{\sqrt{3}}\sigma(w_0,y)\
 .
\end{equation}
This is the general expression for the time drift of the redshift
as it would be measured by an observer at the center of a
spherically symmetric universe.

{\bf Robertson-Walker case.} Since $\sigma=0$ for a RW spacetime,
Eq.~(\ref{gen8}) reduces to the standard Sandage-McVittie formula.
Let us consider the RW metric in conformal coordinates
\begin{equation}\label{FLmetric}
 \dd s^2 =a^2(\eta)\left[
 -\dd\eta^2+\dd\chi^2+f_K^2(\chi)\dd\Omega^2
 \right]\ ,
\end{equation}
where $\chi$ is the radial comoving coordinate and
$f_K=(\sin\chi,\chi,\sinh\chi)$ according to the curvature of the
spatial sections. Now, with $ w= \eta + \chi$ and $y = \chi$, this
leads to the form~(\ref{Obsmetric}) for the
metric~(\ref{FLmetric}) with $A=B=a(w-y)$ and $C= a(w-y)f_K(y)$.
$H$ is thus constant on each constant time hypersurface and
$\sigma=0$ everywhere. Since $\partial_w A|_{y={\rm const.}} =
\partial_\eta a$, Eq.~(\ref{defzdot}) gives $\dot z = (1+z) H_0 -
H(z)$ where we have shifted to cosmic time ($\dd t = a\dd\eta$),
using $1+z=a_0/a$.

We understand from this exercise why the observational coordinates
are adapted to the computation of $\dot z$ in an arbitrary
spacetime. They are just a generalisation to less symmetric
spacetimes of the conformal coordinates, where  the expression is
easily obtained in the RW case.

We also conclude that since the 3 functions $(A,B,C)$ are
expressed in terms of a unique function, all background
observations depend on some function of $H$, $\dot z$ being no
exception.

{\bf Consistency relation.} It follows that, in a RW universe, we
can determine a consistency relation between several observables.
From the metric~(\ref{FLmetric}) and the relation for $\chi(z)$
that follows, one deduces that $H^{-1}(z)= D'(z)\left[ 1 +
\Omega_{K0}H_0^2D^2(z) \right]^{-1/2}$, where a prime stands for
$\partial_z$ and $D(z)=D_L(z)/(1+z)$. This relation is the basis
of the test of the RW structure, as recently proposed in
Ref.~\cite{testCP3}, arguing there that knowledge of $H(z)$ at
different redshifts, from e.g. baryon acoustic oscillations or
differential age estimates of passively evolving galaxies, could
then be used to check this yields the same value of $\Omega_{K0}$.
Here, we argue that it can be implemented using $\dot z(z)$ as an
observational input. Defining
\begin{eqnarray}\label{Crelat}
 {\rm Cop}[D_L(z),\dot z(z),z]&\equiv&
 1
 + \Omega_{K0}H_0^2D(z)^2\nonumber\\&-&
 \left[H_0(1+z) -\dot z(z)
 \right]^2\left[D'(z)\right]^2,
 \end{eqnarray}
we must have ${\rm Cop}[D_L(z),\dot z(z),z]=0$ whatever the matter
content of the universe and the field equations, since it derives
from a purely kinematical relation that does not rely on the dynamics
(i.e. the Friedmann equations). This is a {\it consistency relation}
between independent observables that holds in {\it any}
Robertson-Walker spacetime.

{\bf Spherically symmetric spacetimes.} Writing the LTB metric in
observational coordinates requires the solution of the null
geodesic equation, which is in general possible only numerically.
Consider an LTB spacetime with metric
$$
 \dd s^2 =-\dd t^2 + S^2(r,t)\dd r^2 + R^2(r,t)\dd\Omega^2
$$
where $S(r,t)= R'/\sqrt{1+2E(r)}$ and $\dot
R^2=2M(r)/R(r,t)+2E(r)$, using a dot and prime to refer to
derivatives with respect to $t$ and $r$. The Einstein equations
can be solved parametrically as
$\{R(r,\eta),t(\eta,r)\}=\{\frac{M(r)}{{\cal
E}(r)}\Phi'(\eta),T_0(r)+\frac{M(r)}{[{\cal
E}(r)]^{3/2}}\,\Phi(\eta) \}$ where $\Phi$ is defined by
$\Phi(\eta)=\left(\sinh\eta -\eta,\eta^3/6,\eta-\sin\eta\right)$,
and ${\cal E}(r)= (2E,2,-2E)$ according to whether $E$ is
positive, null or negative.

This solution depends on 3 arbitrary functions of $r$ only,
$E(r)$, $M(r)$ and $T_0(r)$. Their choice determines the model
completely. For instance $(E,M,T_0)=(-K_0r^2,M_0r^3,0)$
corresponds to a RW universe. One can further use the freedom in
the radial coordinate to fix one of the three functions at will so
that one effectively has only 2 arbitrary independent functions.
Assume we fix $M(r)$. We want to determine $\{E(r),T_0(r)\}$ to
reproduce some observables on our past light cone. This can be
represented parametricaly as $\{r(z),E(z),T_0(z)\}$.

Let us sketch the reconstruction and use $r$ as the integration
coordinate, instead of $z$. Our past light cone is defined as
$t=\hat t(r)$ and we set ${\cal R}(r)=R[\hat t(r),r]$. The time
derivative of $R$ is given by $\dot R[\hat t(r),r]\equiv{\cal
R}_1=\sqrt{2M_0r^3/{\cal R}(r) +2E(r)}$. Then we get $R'[\hat
t(r),r]\equiv{\cal R}_2(r) = -[{\cal R}(r) -3(\hat t(r)
-T_0(r)){\cal R}_1(r)/2]E'/E - {\cal R}_1(r)T_0'(r) +{\cal
R}(r)/r$. Finally, more algebra leads to $\dot R'[\hat
t(r),r]\equiv{\cal R}_3(r)=[{\cal R}_1(r) -3 M_0r^3(\hat t(r)
-T_0(r))/{\cal R}^2(r)]E'(r)/2E(r)+M_0r^3T_0'(r)/{\cal R}^2 +{\cal
R}_1(r)/r$. Thus, $\dot R$, $R'$ and $\dot R'$ evaluated on the
light cone are just functions of ${\cal R}(r)$, $E(r)$, $T_0(r)$
and their first derivatives. Now, the null geodesic equation gives
that
$$
 \frac{\dd\hat t}{\dd r}= -\frac{{\cal
 R}_2(r)}{\sqrt{1+2E(r)}},\quad
 \frac{\dd z}{\dd r}=\frac{1+z}{\sqrt{1+2E(r)}}{\cal R}_3(r),
$$
and
$$
\frac{\dd {\cal R}}{\dd r}=\left[ 1 - \frac{{\cal
R}_1(r)}{\sqrt{1+2E(r)}}\right]
 {\cal R}_2(r).
$$
These are 3 first order differential equations relating 5
functions ${\cal R}(r)$, $\hat t(r)$, $z(r)$ $E(r)$ and $T_0(r)$.
To reconstruct the free functions we thus need 2 observational
relations. ${\cal R}(z)=D_A(z)$ is the obvious choice. Then, from
$\dot z(z)$, we have the new relation
\begin{equation}\label{newcond}
 \frac{1}{9}\left(\frac{{\cal R}_3}{{\cal R}_2}-\frac{{\cal R}_1}{{\cal
 R}}\right)^2 =
 \left[\dot z - (1+z)H_0
 + \frac{1}{3}\left(\frac{{\cal R}_3}{{\cal R}_2}+2\frac{{\cal R}_1}{{\cal
 R}}\right)\right]^2.
\end{equation}
It was shown~\cite{ltb1} that the observed $D_L(z)$ can be
reproduced from the function $T_0(r)$ assuming $E=0$, or from the
function $E(r)$ assuming $T_0=0$. Here, we have shown that
$\{E(r),T_0(r)\}$ can be completely reconstructed from the data
without assumptions.

\section{Discussion}

\vskip-.3cm

In this letter, we have shown that observation of both the
luminosity distance and time drift of the redshift as a function
of $z$ allows one to construct a test of the Copernican principle.
We have derived the general expression for $\dot z(z)$ in a
spherically symmetric spacetime [see Eqs.~(\ref{defzdot})
and~(\ref{gen8})]. This extends the standard computation which was
restricted to RW spacetimes and was extended to almost RW
spacetimes only recently~\cite{deltazdot}.

As a byproduct, this also allowed us to derive the consistency
relation~(\ref{Crelat}) between the observed $D_L(z)$ and $\dot
z(z)$, thereby extending the result of Ref.~\cite{testCP3} to an
alternate observable. That is, while $H(z)$ characterises the
local isotropic expansion rate, $\dot z(z)$ gives access to the
expansion between us and a source. In RW models these are
trivially related but in general the shear enters these
observables differently thereby presenting a test of the
Copernican principle using only background observations. We
have shown that we can extract the shear as a function of $z$ and
demonstrated that it allows one to close the reconstruction
problem for a LTB spacetime.

$D_L(z)$ can be measured from the observation of Type~Ia
supernovae, particularly with actual projects such as JDEM, up to
redshifts of a few. $\dot z(z)$ has a typical amplitude of order
$\delta z\sim -5\times10^{-10}$ on a time scale of $\delta t=
10$~yr, for  a source at redshift $z=4$. This measurement is
challenging, and impossible with present-day facilities. However,
it was recently revisited~\cite{Loeb} in the context of ELT,
arguing they could measure velocity shifts of order $\delta v\sim
1-10\ {\rm cm/s}$ over a 10 year period from the observation of
the Lyman-$\alpha$ forest. It is one of the science drivers in
design of the CODEX spectrograph~\cite{Pasquini1} for the future
European ELT. The study of the precision to which we can check the
Copernican principle with these two data sets is beyond the scope
of this letter. Indeed, many effects, such as proper motion of the
sources, local gravitational potential, or acceleration of the Sun
may contribute to the time drift of the redshift. It was
shown~\cite{deltazdot}, however, that these contributions can be
brought to a 0.1\% level so that the cosmological redshift is
actually measured.

Future high precision data may thus allow a test of the Copernican
principle, even though observations are localized on our past
light cone. While important in its own right for understanding the
foundations of our cosmological model, it is also critical for our
understanding of the acceleration of the universe~-- it will
permit us to be confident that any such acceleration is not simply
a misinterpretation of the data because of incorrectly assuming
the geometry of our universe at low redshift.

\noindent{\bf Acknowledgements}: We thank B. Bassett, F. Bernardeau, Y. Mellier
for discussions. CAC is funded by the NRF (South Africa).


\end{document}